\documentclass[preprint,showpacs,showkeys,preprintnumbers,amsmath,amssymb]{revtex4} 

\usepackage{graphicx}
\usepackage{dcolumn}
\usepackage{bm}

\begin{document}


\title{Solution of reduced equations derived with singular
perturbation methods }

\author{Masatomo Iwasa}
\affiliation{%
Department of Physics, Nagoya University,\\
Nagoya 464-8602, Japan\\
}%


\date{\today}

\begin{abstract}
For singular perturbation problems in dynamical systems, various
 appropriate singular perturbation methods have been proposed to eliminate
 secular terms appearing in the naive expansion.  
For example, the method of multiple time scales, the normal form method, center
 manifold theory, the renormalization group method are well known.  
In this paper, it is shown that all of the solutions of the reduced
 equations constructed with those methods are exactly equal to sum of the
 most divergent secular terms appearing in the naive expansion. 
For the proof, a method to construct a perturbation solution which
 differs from the conventional one is presented,
 where we make use of the theory of Lie symmetry group.   
\end{abstract}

\maketitle
\section{introduction}
This paper investigates perturbation analysis of the fundamental
system of nonlinear ordinary differential equations,
\begin{eqnarray}
   \frac{du}{dt}:=\dot{u}=Mu+\varepsilon f(u), 
\label{eq:1-1}
\end{eqnarray}
where $u\in \mathbb{C}^n$ is the dependent variable, $M$ is a constant
$n\times n$ matrix,
$f:\mathbb{C}^n\rightarrow\mathbb{C}^n$ is nonlinear function of $u$,
and $\varepsilon\in\mathbb{R}$ is a perturbation parameter. 
In what follows, we refer to $\dot{u}=Mu$ as the unperturbed system and 
(\ref{eq:1-1}) as the perturbed system. 
The simplest perturbation solution is the naive expansion.
Let us pose an expansion for the solution in powers of $\varepsilon$, 
\begin{eqnarray}
  u(t;\varepsilon) = \sum_{p=0}^{\infty}\varepsilon^pu^{(p)}(t).
\label{eq:1-2}
\end{eqnarray}
If we substitute Eq. (\ref{eq:1-2}) into Eq. (\ref{eq:1-1}), expand the
both side of the equation with respect to $\varepsilon$ and equate the
coefficients of each power of $\varepsilon$, 
then we obtain the following series of differential equations: 
\begin{eqnarray}
  &&\dot {u}^{(0)}=Mu^{(0)},
   \nonumber \\
  &&\dot {u}^{(p)}=Mu^{(p)}+f^{(p-1)}(u^{(0)},u^{(1)},\ldots,u^{(p-1)})
   \ \ {\rm for}\ \ p=1, 2, \ldots ,  
\label{eq:1-3}
\end{eqnarray} 
where 
\begin{eqnarray}
  f\left(\sum_{p=0}^{\infty}\varepsilon^pu^{(p)}\right)
    =:\sum_{p=0}^{\infty}\varepsilon^pf^{(p)}
    \left(u^{(0)},u^{(1)},\cdots,u^{(p)}\right).  
\label{eq:1-4}
\end{eqnarray}
If we solve these equations recursively, 
we find the naive expansion.

In this paper, we are especially interested in singular perturbation
problems where secular terms arise in the naive expansion.  
In general, if $f(u)$ is a power series, secular terms arise in the
naive expansion as we see later.  
To eliminate those secular terms, various appropriate methods are
developed such as, for example, 
the renormalization group method \cite{ChenGoldenfeldOono1996,
GotoMasutomiNozaki1999, Chiba20081, Chiba20082, Chiba20083,
IwasaNozaki2006},    
the method of multiple time scales,\cite{Nayfeh},
canonical Hamiltonian perturbation theory \cite{Hori1966},
the averaging methods\cite{BogolyubovPlace}, the method of normal forms
\cite{Nayfeh},  
center manifold theory \cite{GuckenheimerHolmes}, and so on. 
We refer to these methods simply as singular perturbation methods in
this paper.  
It is well known that all of these methods result in equations all of
which are equivalent that govern the long-time behavior of the system.
Each of them is the dynamics for integral constants of unperturbed
system, or in other words, dynamics in the null space of the linear
operator determined from the unperturbed system.
Although the name of that equation such as the renormalization group
equation or the normal form depends on the method,
in this paper, we refer to it simply as reduced equation.  

There have been many studies which show those singular
perturbation methods surely lead to the well-behaving approximate
solution.
However, what the solution of that reduced equation exactly includes has
not been clear.   
In the paper, we reveal the exact solution of the reduced equation. 
To be precise, the following statement is the main result shown in this
paper.\\ 
{\it Main result : 
the solution of the reduced equation up to first order for singular
perturbation problem (\ref{eq:1-1}) is equal to sum of those terms which
are proportional to $\varepsilon t,\ \varepsilon^2t^2,\
\varepsilon^3t^3,\ldots$ in the naive expansion . \\
}
In what follows, we refer to those secular terms as most divergent
terms in the naive expansion.
Although this fact has been believed to be true in some cases
\cite{Goldenfeld}, 
this is rigorously proved in this paper.

In the proof of the result, we first present another method to
construct a perturbation solution, that is in Proposition 1 in section
II. 
While $f^{(p)}$ in Eqs. (\ref{eq:1-3}) must be more complicated function
of $u^{(0)}, u^{(1)}, \ldots, u^{(p)}$ as $p$ becomes large in
general.  
the method presented in section II leads to another recursive equations
which has clearer expression compared with the Eqs. (\ref{eq:1-3}).
In the derivation of those recursive equations, we make use of Lie
symmetry group which leaves the system Eq. (\ref{eq:1-1}) invariant.
This method can be interpreted as a extension of the renormalization
group method with Lie symmetry group \cite{IwasaNozaki2006}.
The recursive equation plays an important role in section III for the
proof of the main result.

\section{A method to construct a perturbation solution}
Consider the system of nonlinear ordinary differential equations as
follows: 
\begin{eqnarray}
 \frac{du}{dt}:=\dot {u}=Mu+\varepsilon f(u), 
\label{eq:2-1}
\end{eqnarray}
where $u=u(t)\in \mathbb{C}^n$ is a vector valued function of an
independent valuable, $M$ is an $n \times n$ 
matrix whose all coefficients are constant, $\varepsilon$ is a constant
and $f :\mathbb{C}^ n \rightarrow\mathbb{C}^n$ is a smooth vector valued
function. 
In what follows, we refer to $\dot{u}=Mu$ as
the unperturbed system and (\ref{eq:2-1}) as the perturbed system, and
the solution of the system (\ref{eq:2-1}) is denoted by
$u=u(t;\varepsilon)$ for the dependence to the perturbation parameter
$\varepsilon$. 

Firstly let us find a method to construct a perturbation solution.
For the construction, we make use of the Lie symmetry method
\cite{Olver}.

{\bf Proposition. 1}\ \ {\it
Suppose $\psi(t,u;\varepsilon)\in\mathbb{C}^n$ is a vector valued
function of $t$, $u$ and $\varepsilon$, and 
its formal expansion in powers of $\varepsilon$,
$  \psi(t,u;\varepsilon)=:\sum_{r=0}^{\infty}\varepsilon^r\psi^{(r)}(t,u),
$
is admitted.
Then, for $\psi(t,u;\varepsilon)$ which satisfies the recursive differential
equations, 
\begin{eqnarray}
  L\psi^{(0)}&\hspace{-0.2cm}=&\hspace{-0.2cm}f,
   \nonumber \\
  L\psi^{(r)}&\hspace{-0.2cm}=&\hspace{-0.2cm}
                          \psi^{(r-1)}\cdot\partial_{u}f 
                          -f\cdot\partial_{u}\psi^{(r-1)}\ \ \ 
  \ \ {\rm for}\ \ r=1,2,3,\ldots ,
\label{eq:2-4}\\
   L&:=&I\left(\partial_t+(Mu)\cdot\partial_{u}\right)-M,
  \nonumber 
\end{eqnarray}
the solution of system (\ref{eq:2-1}) satisfies,
\begin{eqnarray}
   u(t;\varepsilon)
     &=&u(t;0)+\sum_{r=0}^{\infty}\int_{0}^{\varepsilon}
                \sigma^r\psi^{(r)}(t,u(t;\sigma))
               d\sigma. 
\label{eq:2-3}
\end{eqnarray}
Here $I$ in the definition of $L$ denotes the identity matrix.
}

{\bf Proof.}\ \ 
Suppose Eq. (\ref{eq:2-1}) admits a Lie symmetry group whose
infinitesimal generator is denoted by 
\begin{eqnarray}
  X:=\partial_\varepsilon+\psi(t,u;\varepsilon)\cdot\partial_{u}.
\label{eq:2-6}
\end{eqnarray}  
Then its prolongation $X^*$, 
\begin{eqnarray}
  X^*&=&\partial_\varepsilon+\psi(t,u;\varepsilon)\cdot\partial_{u}
       +\psi^{\dot{u}}(t,u,\dot{u};\varepsilon)\cdot\partial_{\dot{u}},
\label{eq:2-7}\\
  \psi^{\dot{u}}(t,u,\dot{u};\varepsilon)&:=&\left[\partial_t
+\dot{u}\cdot\partial_{u}\right]\psi(t,u;\varepsilon),
\nonumber
\end{eqnarray}   
satisfies the infinitesimal criterion of invariance of system
(\ref{eq:2-1}), that is 
\begin{eqnarray}
  X^*\left[\dot{u}-Mu-\varepsilon f(u)\right]
      \bigr|_{{\rm Eq.(\ref{eq:2-1})}}
    &=&0.
\label{eq:2-8}
\end{eqnarray}
Eq. (\ref{eq:2-8}) reads
\begin{eqnarray}
  \left[
     I\left(
       \partial_t+(Mu)\cdot\partial_{u}
      \right)
     \psi
     -M\psi-f
  \right] 
  +\varepsilon\left[
                 f\cdot\partial_{u}\psi
                 -\psi\cdot\partial_{u}f
              \right]     
    =0. 
\label{eq:2-9}
\end{eqnarray}
For the formal expansion in powers of $\varepsilon$,
\begin{eqnarray}
  \psi(t,u;\varepsilon)=\sum_{r=0}^{\infty}\varepsilon^r\psi^{(r)}(t,u),
\label{eq:2-10}
\end{eqnarray}
by substituting Eq. (\ref{eq:2-10}) into Eq. (\ref{eq:2-9}) and equating
the coefficient of each $\varepsilon^r$, we find recursive equations as
follows:   
\begin{eqnarray}
 \hspace{-0.2cm}
  L\psi^{(0)}&\hspace{-0.2cm}=&\hspace{-0.2cm}f,
\label{eq:2-11-1}   \\
 \hspace{-0.2cm}
  L\psi^{(r)}
     &\hspace{-0.2cm}=&\hspace{-0.2cm}
          \Bigl[
             \psi^{(r-1)}\cdot\partial_{u}f
             -f\cdot\partial_{u}\psi^{(r-1)}
          \Bigr],\ \ \ {\rm for} \  
      r=1, 2,\ldots,  
\label{eq:2-11-2}  \\
   L&:=&I
              \left(
                 \partial_t+(Mu)\cdot\partial_{u}
              \right)
             -M. 
\nonumber
\end{eqnarray}
Solving Eq. (\ref{eq:2-11-1}) and Eq. (\ref{eq:2-11-2}) recursively, we obtain
formal expansion of the infinitesimal generator of a Lie symmetry group
which leaves system (\ref{eq:2-1}) invariant. 
Then the solution of system (\ref{eq:2-1}), $u=u(t;\varepsilon)$,
invariant to $X$ satisfies
\begin{eqnarray}
  X\left[u-u(t;\varepsilon)\right]\bigr|_{u=u(t;\varepsilon)}=0.
\label{eq:2-13}
\end{eqnarray}
Eq. (\ref{eq:2-13}) reads
\begin{eqnarray}
  \frac{\partial}{\partial \varepsilon}u(t;\varepsilon)
     &=&\psi(t,u(t;\varepsilon);\varepsilon) 
\label{eq:2-14.5} \\
  \Leftrightarrow
  u(t;\varepsilon)
     &=&u(t;0)+\sum_{r=0}^{\infty}\int_{0}^{\varepsilon}
                \sigma^r\psi^{(r)}(t,u(t;\sigma))
               d\sigma. 
\label{eq:2-14}
\end{eqnarray}
Thus, the integral equation (\ref{eq:2-3}) for the solution of system
(\ref{eq:2-1}) has been obtained. $\Box$ 

It should be remarked here that it is not necessary for $\varepsilon$ to
be small in this proposition.
Therefore, Eq. (\ref{eq:2-14.5}) holds not only for perturbed systems but
also for generic systems which take the form of Eq. (\ref{eq:2-1})
although it seems to be practical for perturbation problems.

\section{The solution of reduced equations}
Next, consider those systems whose linear part can be diagonalized.
Then the system (\ref{eq:2-1}) reads
\begin{eqnarray}
  \dot{z} = \Lambda z+\varepsilon g(z), \ \ \ z\in\mathbb{C}^n,
\label{eq:2-15}
\end{eqnarray} 
with a linear transformation from $u$ into $z$.
Here $\Lambda$ is an $n \times n$ 
diagonal matrix whose components are denoted by
$\Lambda_{ij}=:\delta_{ij}\lambda_i$,
and $g$ is the nonlinear vector valued function constructed from $f$
with the transformation.   
Then the recursive equations corresponding to Eqs. (\ref{eq:2-4}) become 
\begin{eqnarray}
  L_i\phi^{(0)}_i&=&g_i,
   \nonumber \\
  L_i\phi^{(r)}_i&=&\sum_{j=1}^{n}
                      \Bigl[
                        \phi^{(r-1)}_j\partial_{z_j}g_i
                        -g_j\partial_{z_j}\phi^{(r-1)}_i
                      \Bigr]\ \ {\rm for}\ r=1,2,\ldots,   
   \label{eq:2-18}\\ 
 L_i&:=&\left(\partial_t+\sum_{k=1}^{n}\lambda_k z_k\partial_{z_k}\right)-\lambda_i,\nonumber 
\end{eqnarray}
for a vector valued function, $\phi(t,z;\varepsilon)\in\mathbb{C}^n$ for
$r=0,1,\ldots$.  
Here and in what follows, the components of vectors and matrices are
explicitly denoted for the clarification of the following discussion,
and equations hold for $i=1,\ldots,n$. 
In the same way we have derived Eq. (\ref{eq:2-3}), it follows that the
solution of Eq. (\ref{eq:2-15}), $z=z(t;\varepsilon)$, satisfies 
\begin{eqnarray}
  z_i(t;\varepsilon)=z_i(t;0)+\sum_{r=0}^{\infty}\int_{0}^{\varepsilon}\varepsilon^r\phi_i^{(r)}(t,z(t;\varepsilon))d\varepsilon.
\label{eq:2-16}
\end{eqnarray}
As we see later, if we obtain $\{\phi^{(r)}(t,z)\}$,
we can construct the naive expansion using Eq. (\ref{eq:2-16}) with
iterative method since we know the solution of the unperturbed system, 
$z(t;0)={\rm e}^{\Lambda t}z_0$ where $z_0$ is a constant. 

Now we can show the following proposition.

{\bf Proposition. 2}\ \ {\it 
Suppose the nonlinear function in Eq. (\ref{eq:2-15}) is power series
such as
\begin{eqnarray}
 g_i(z)=\sum_{p_1,p_2,\ldots,p_n=0}^{\infty}C^i_{p_1 p_2 \ldots p_n}
         \prod_{k=1}^n z_k^{p_k},
 \label{eq:3-1}
\end{eqnarray}
where each $C^i_{p_1 p_2 \ldots p_n}\ $ is constant.
Then there is a solution of Eqs. (\ref{eq:2-18}) which
becomes power series of $t$ and $z$ which 
satisfies $\phi_i^{(r)} =  O(t^r)$ for $r=1,2,\ldots$ 
while $\phi_i^{(0)}=O(t)$.}

{\bf Proof.}\ \ 
Firstly, we seek $\phi^{(0)}$. 
According to Eqs. (\ref{eq:2-18}), it is the solution of the
differential equation, 
\begin{eqnarray}
 L_i \phi^{(0)}_i(t,z)=\sum_{p_1,p_2,\ldots,p_n=0}^\infty C^i_{p_1 p_2 \ldots
  p_n}\prod_{k=1}^n z_k^{p_k}.
 \label{eq:3-2}
\end{eqnarray}
Note that, for arbitrary $(p_1,\ p_2,\ \ldots,p_n )\in\mathbb{N}^n$,
$\prod_{k=1}^n z_k^{p_k}$ are eigenfunctions of $L_i$. 
Among inhomogeneous terms in the right hand side,
those which satisfy the resonance condition,
 $\sum_{j=1}^n\lambda_jp_j-\lambda_i=0$,  
cause secular terms in the solution.
Then we obtain,
\begin{eqnarray}
 \phi^{(0)}_i(t,z)&=&\hspace{-0.5cm}
               \sum_{\scriptsize{
               \begin{array}{c}
		p_1,p_2,\ldots,p_n\\
                \sum_{j=1}^n \lambda_jp_j-\lambda_i=0
	       \end{array}}
               }
                   \hspace{-0.5cm}
           C^i_{p_1 p_2 \ldots p_n}t\prod_{j=1}^n z_j^{p_j}
          + \hspace{-0.5cm}
              \sum_{\scriptsize{
               \begin{array}{c}
		p_1,p_2,\ldots,p_n\\
                \sum_{j=1}^n \lambda_jp_j-\lambda_i\neq 0
	       \end{array}}
               }
                   \hspace{-0.7cm}
           \frac{C^i_{p_1 p_2 \ldots p_n}}{\sum_{j=1}^n\lambda_jp_j-\lambda_i}
           \prod_{j=1}^n z_j^{p_j}.
 \label{eq:3-3}
\end{eqnarray}
Next, we seek $\phi^{(1)}$.
According to Eqs. (\ref{eq:2-18}), it is the solution of the
differential equation,
\begin{eqnarray}
 L_i\phi^{(1)}_i=\sum_{j=0}^{n}
                  \left[\phi^{(0)}_j\partial_{z_j}g_i
                 -g_j\partial_{z_j}\phi^{(0)}_i\right].
 \label{eq:3-4}
\end{eqnarray} 
By virtue of $\phi^{(0)}$, the inhomogeneous terms in Eq. (\ref{eq:3-4})
can be split into four parts as 
\begin{eqnarray}
  \left[{\rm r. h. s.\ of\ Eq. (\ref{eq:3-4})}\right]
   =           \hspace{-0.5cm}
               \sum_{\scriptsize{
               \begin{array}{c}
		p_1,p_2,\ldots,p_n\\
                \sum_{j=1}^n \lambda_jp_j-\lambda_i=0
	       \end{array}}
               }
                   \hspace{-0.5cm}
           E^i_{p_1 p_2 \ldots p_n}t\prod_{j=1}^n z_j^{p_j}
     +         \hspace{-0.5cm}
               \sum_{\scriptsize{
               \begin{array}{c}
		p_1,p_2,\ldots,p_n\\
                \sum_{j=1}^n \lambda_jp_j-\lambda_i\neq0
	       \end{array}}
               }
                   \hspace{-0.5cm}
           F^i_{p_1 p_2 \ldots p_n}t\prod_{j=1}^n z_j^{p_j}
       \nonumber \\
     +         \hspace{-0.5cm}
               \sum_{\scriptsize{
               \begin{array}{c}
		p_1,p_2,\ldots,p_n\\
                \sum_{j=1}^n \lambda_jp_j-\lambda_i=0
	       \end{array}}
               }
                   \hspace{-0.5cm}
           G^i_{p_1 p_2 \ldots p_n}\prod_{j=1}^n z_j^{p_j}
     +         \hspace{-0.5cm}
               \sum_{\scriptsize{
               \begin{array}{c}
		p_1,p_2,\ldots,p_n\\
                \sum_{j=1}^n \lambda_jp_j-\lambda_i\neq0
	       \end{array}}
               }
                   \hspace{-0.5cm}
           H^i_{p_1 p_2 \ldots p_n}\prod_{j=1}^n z_j^{p_j},
 \label{eq:3-5}
\end{eqnarray}
for some constants 
$\{E^i_{p_1 p_2 \ldots p_n},\ F^i_{p_1 p_2 \ldots p_n},\ 
G^i_{p_1 p_2 \ldots p_n}\ $, $H^i_{p_1 p_2 \ldots p_n}\}$.
All terms in the first part seem to cause secular terms in $\phi^{(1)}$
which are proportional to $t^2$ since each of them satisfies resonance
condition. 
However, we can show the first part vanishes 
by substituting Eq. (\ref{eq:3-1}) and Eq. (\ref{eq:3-3})
into the right hand side of Eq. (\ref{eq:3-4}) and calculating
$\{E^i_{p_1p_2\ldots p_n}\}$. 
The calculation is found in Appendix A concretely. 
Therefore, the most divergent terms in
$\phi^{(1)}$ are not proportional to $t^2$ but proportional to $t$.
For $r=2,3,\ldots$, 
inhomogeneous terms in Eq. (\ref{eq:2-18}) which are proportional to
$t^{r-1}$ and which satisfy the resonance condition remain in general.  
Then those inhomogeneous terms cause secular terms proportional to
$t^{r}$ in $\phi^{(r)}$. 
$\Box$

Now we can find the solution of reduced equations which result
from various singular perturbation methods.

{\bf Corollary.}\ \ {\it
For the system of differential equations, 
 \begin{eqnarray}
   \frac{\partial z(t;\varepsilon)}{\partial\varepsilon} =
    t\phi^{(0)}_{sec}(z(t;\varepsilon)), 
 \label{eq:3-6}
 \end{eqnarray}
with $z(t;0)$ denoting the solution of the unperturbed system,
the solution is equal to sum of terms proportional to  
$\varepsilon t,\ \varepsilon^2t^2,\ldots,\varepsilon^nt^n,\ldots$
in the naive expansion of system (\ref{eq:2-15}).
Here $\phi^{(0)}$ is split into
$\phi^{(0)}(t,z)=:t\phi_{sec}^{(0)}(z)+\phi_{non}^{(0)}(z)$
by virtue of Eq. (\ref{eq:3-3}). 
}

{\bf Proof.}\ \ 
According to Eq. (\ref{eq:2-16}),
\begin{eqnarray}
 z(t;\varepsilon)=z(t;0)&+&
         \int_{0}^{\varepsilon}t\phi^{(0)}_{sec}(z(t;\sigma))d\sigma
        +\int_{0}^{\varepsilon} \phi^{(0)}_{non}(z(t;\sigma))d\sigma
                   \nonumber \\
&& +\int_{0}^{\varepsilon}\varepsilon\phi^{(1)}(t,z(t;\sigma))d\sigma
 +\int_{0}^{\varepsilon}\varepsilon^2\phi^{(2)}(t,z(t;\sigma))d\sigma
               +\cdots.
 \label{eq:3-8}
\end{eqnarray}
Thanks to this self-consistent integral equation, 
we can construct the naive expansion with iterative method.
In terms of Proposition 2,
it follows that terms proportional to 
$\varepsilon t, \varepsilon^2t^2,\ldots, \varepsilon^nt^n,\ldots$ 
in the naive expansion arise only from the term
$\int_{0}^{\varepsilon}t\phi^{(0)}_{sec}(t;\sigma)d\sigma$
among terms in the right hand side of Eq. (\ref{eq:3-8}) any step of the
iteration.
Therefore, the solution of the following equation (\ref{eq:3-10}) is
exactly equal to sum of terms proportional to  
$\varepsilon t,\ \varepsilon^2t^2,\ldots,\ \varepsilon^nt^n,\ldots$ 
in the naive expansion;
\begin{eqnarray}
  &&z(t;\varepsilon)=z(t;0)+\int_{0}^{\varepsilon}t\phi^{(0)}_{sec}(z(t;\sigma))d\sigma,
 \label{eq:3-9.5}\\
  &&\Leftrightarrow
  \frac{\partial z(t;\varepsilon)}{\partial\varepsilon}=t\phi^{(0)}_{sec}(z(t;\varepsilon)),
 \label{eq:3-10}
\end{eqnarray}
where we adopt the solution of the unperturbed system as $z(t,0)$.
$\Box$

To complete the proof of the main result, we have to show various
 the widely-accepted reduced equations is equivalent to
 Eq. (\ref{eq:3-6}). 
As a result of singular perturbation methods, we obtain reduced
equations such as 
\begin{eqnarray}
  \frac{\partial z(t;\varepsilon)}{\partial t}=\varepsilon
   z^{(1)}_{sec}(z(t;\varepsilon)),  
 \label{eq:3-11.5}
\end{eqnarray}
where $z^{(1)}$ denotes the coefficient of $\varepsilon$ in the naive
expansion and we set
$z^{(1)}(t,z^{(0)})=:tz^{(1)}_{sec}(z^{(0)})+z^{(1)}_{non}(z^{(0)})$. 
Eq. (\ref{eq:3-11.5}) is a normal form expression of the reduced
equations. Although the well-known normal form contains linear part 
\cite{Nayfeh} such as   
\begin{eqnarray}
 \frac{\partial \tilde{z}(t;\varepsilon)}{\partial t}
   = \Lambda \tilde{z}(t;\varepsilon) 
     +\varepsilon\phi^{(0)}_{sec}(\tilde{z}(t;\varepsilon)),
 \label{eq:3-14} 
\end{eqnarray} 
Eq. (\ref{eq:3-14}) reads Eq. (\ref{eq:3-12}) under
$z:=\exp{(-\Lambda t)}\tilde{z}$. 
We can transform Eq. (\ref{eq:3-12}) into renormalization group equation
or equivalent reduced equations derived with other methods
if we adopt integral constants appearing in the solution of the
unperturbed system as dependent variables \cite{ChenGoldenfeldOono1996}. 
The equivalence of the normal form theory and the renormalization group
method is discussed in \cite{DeVille2007} in detail.  
 
Eq. (\ref{eq:3-11.5}) reads
\begin{eqnarray}
\frac{\partial z(t;\varepsilon)}{\partial
 t}=\varepsilon\phi^{(0)}_{sec}(z(t;\varepsilon)) 
 \label{eq:3-12} 
\end{eqnarray}
for the following reason: 
For the expanded form of the solution,
$z(t;\varepsilon)=:\Sigma_{k=0}^{\infty}\varepsilon^kz^{(k)}(t)$,
$z^{(1)}$ satisfies 
\begin{eqnarray}
&&\hspace{-0.3cm} \dot{z^{(1)}_i}(t)=\lambda_i z^{(1)}_i(t)+ g_i(z^{(0)}(t))
  \nonumber \\
&&\hspace{-0.3cm} \Leftrightarrow\left(\partial_t+\sum_{k=1}^{n}\dot{z_k}^{(0)}\partial_{z^{(0)}_k}\right)z^{(1)}_i(t,z^{(0)})=\lambda_i z^{(1)}_i(t,z^{(0)}) + g_i(z^{(0)})
  \nonumber \\
&&\hspace{-0.3cm} \Leftrightarrow\left(\partial_t+\sum_{k=1}^{n}\lambda_kz_k^{(0)}\partial_{z^{(0)}_k}-\lambda_i\right)z^{(1)}_i(t,z^{(0)})= g_i(z^{(0)}).
 \label{eq:3-11}
\end{eqnarray}
Eq. (\ref{eq:3-11}) corresponds to the first equation of (\ref{eq:2-18})
if we replace $z^{(1)}$ and $z^{(0)}$ to $\phi^{(0)}$ and $z$
respectively. 
With a new independent valuable $\tau:=\varepsilon t$,
both Eq. (\ref{eq:3-6}) and Eq. (\ref{eq:3-12}) can be written as 
\begin{eqnarray}
 \frac{d z(\tau)}{d \tau}=\phi^{(0)}_{sec}(z(\tau)).
 \label{eq:3-13} 
\end{eqnarray}
Thus, we have shown that the solution of the reduced
equations is equal to sum of the most divergent terms in the naive 
expansion when we construct the reduced equations up to only first
order.  

\section{Example: The Duffing Equation}
Let us see what is shown above holds through a simple example.  
Consider the Duffing equation,
\begin{eqnarray}
 \ddot{u}+u=\varepsilon u^3.
\label{eq:4-1}
\end{eqnarray}
Introducing $z:=u+{\rm i}\dot{u}$ for simplicity, we have
\begin{eqnarray}
 \dot{z}+{\rm i}z =\varepsilon\frac{{\rm i}}{8}(z+\bar{z})^3.
\label{eq:4-2}
\end{eqnarray}

Firstly, let us review the proof of the main result with this example. 
Suppose Eq. (\ref{eq:4-2}) admits the Lie symmetry group whose
infinitesimal generator is denoted by 
\begin{eqnarray}
 X:=\partial_\varepsilon
     +\psi^z(t,z,\bar{z};\varepsilon)\partial_z
     +\psi^{\bar{z}}(t,z,\bar{z};\varepsilon)\partial_{\bar{z}}.
\label{eq:4-11}
\end{eqnarray}
Note that it can be shown 
$\psi^z(t,z,\bar{z};\varepsilon)
=\overline{\psi^{\bar{z}}(t,z,\bar{z};\varepsilon)}$.  
Then its prolongation $X^*$,
\begin{eqnarray}
 X^*&=&\partial_\varepsilon
     +\psi^z(t,z,\bar{z};\varepsilon)\partial_z
     +\psi^{\bar{z}}(t,z,\bar{z};\varepsilon)\partial_{\bar{z}}
\nonumber \\
  && +\psi^{\dot{z}}(t,z,\bar{z},\dot{z},\dot{\bar{z}};\varepsilon)
       \partial_{\dot{z}}
     +\psi^{\dot{\bar{z}}}(t,z,\bar{z},\dot{z},\dot{\bar{z}};\varepsilon)
          \partial_{\dot{\bar{z}}}, 
\label{eq:4-12}\\
\psi^{\dot{z}}(t,z,\bar{z},\dot{z},\dot{\bar{z}};\varepsilon)
  &:=&\left[\partial_t+\dot{z}\partial_z
          +\dot{\bar{z}}\partial_{\bar{z}}\right]
	  \psi^z(t,z,\bar{z};\varepsilon),
\nonumber \\
  &=&\overline{\psi^{\dot{\bar{z}}}
   (t,z,\bar{z},\dot{z},\dot{\bar{z}};\varepsilon)},  
\nonumber
\end{eqnarray}
satisfies the infinitesimal criterion of the invariance corresponding to
Eq (\ref{eq:2-8}), 
\begin{eqnarray}
 X^*\left[ \dot{z}+{\rm i}z - \varepsilon
           \frac{{\rm i}}{8}(z+\bar{z})^3\right]
     \Biggr|_{{\rm Eq. (\ref{eq:4-2})}}=0. 
 \label{eq:4-13}
\end{eqnarray}
For the formal expanded form of $\psi$
\begin{eqnarray}
 \psi^z(t,z,\bar{z};\varepsilon)
  =\sum_{r=0}^{\infty} \varepsilon^r  \psi^{(r)}(t,z,\bar{z}), 
 \label{eq:4-14}
\end{eqnarray}
the equation for the leading order becomes
\begin{eqnarray}
 (\partial_t-iz\partial_z+i\bar{z}\partial_{\bar{z}}+i)
  \psi^{(0)}(t,z,\bar{z})= \frac{{\rm i}}{8}(z+\bar{z})^3.
 \label{eq:4-15}
\end{eqnarray}
Solving this, we obtain
\begin{eqnarray}
  \psi^{(0)}(t,z,\bar{z})
             =-\frac{1}{16}z^3+\frac{3{\rm i}}{8}t|z|^2z
              +\frac{3}{16}|z|\bar{z}+\frac{1}{32}\bar{z}^3.
 \label{eq:4-16}
\end{eqnarray}
The reduced equation corresponding to Eq. (\ref{eq:3-6}) becomes
\begin{eqnarray}
 \frac{\partial z(t,\varepsilon)}{\partial \varepsilon} 
   = \frac{3{\rm i}}{8}t|z(t;\varepsilon)|^2 z(t;\varepsilon).
 \label{eq:4-17}
\end{eqnarray}
With the the integral equation expression corresponding to
(\ref{eq:3-9.5}), we can find the solution with iterative method. 
The solution up to third order becomes
\begin{eqnarray}
 z(t,\varepsilon) = 
    A{\rm e}^{-{\rm i}t}
    +\frac{3{\rm i}}{8}\varepsilon t |A|^2A{\rm e}^{-{\rm i}t}
    -\frac{9}{128}\varepsilon^2t^2|A|^4A{\rm e}^{-{\rm i}t}
    -\frac{9{\rm i}}{1024} \varepsilon^3t^3|A|^6A{\rm e}^{-{\rm i}t}
    +\cdots,
 \label{eq:4-18}
\end{eqnarray}
where $A$ denotes the integral constant.
We can immediately show this solution is exactly equal to the most
divergent terms in the naive expansion by constructing it directly.

Next, let us see Eq. (\ref{eq:4-17}) is equivalent to reduced equations
derived with the conventional singular perturbation methods. 
Although there are many ways to represent the reduced equations, one of 
them is the normal form \cite{Nayfeh},
\begin{eqnarray}
 \frac{d \tilde{z}(t)}{d t}=-i\tilde{z}(t)
   +\varepsilon\frac{3{\rm i}}{8}|\tilde{z}(t)|^2\tilde{z}(t),
\label{eq:4-6}
\end{eqnarray}
which corresponds to (\ref{eq:3-14}). Under $\tilde{z}=:A(t){\rm
e}^{-it}$, Eq. (\ref{eq:4-6}) reads
\begin{eqnarray}
 \frac{d A(t)}{d t}=\varepsilon\frac{3{\rm i}}{8}|A(t)|^2A(t).
\label{eq:4-7}
\end{eqnarray}
which corresponds to (\ref{eq:3-11.5}).
If we set $A(t)=:R(t){\rm e}^{-{\rm i}\theta(t)}$ where $R(t),\
\theta(t)\in\mathbb{R}$, the reduced equation reads
\begin{eqnarray}
 \frac{d R}{d t}     &=&0, \\
 \frac{d \theta}{d t}&=&-\varepsilon\frac{3}{8}R^2.
\label{eq:4-8}
\end{eqnarray}
which is called renormalization group equation
\cite{ChenGoldenfeldOono1996, GotoMasutomiNozaki1999}. 
Under $\tau:=\varepsilon t$, 
both of Eq. (\ref{eq:4-17}) and Eq. (\ref{eq:4-7}) read
\begin{eqnarray}
 \frac{d\hat{z}(\tau)}{d\tau} 
  = \frac{3{\rm i}}{8}\left|\hat{z}(\tau)\right|^2\hat{z}(\tau).  
 \label{eq:4-18} 
\end{eqnarray} 
Thus, the equivalence has been shown for the Duffing equation

\section{Concluding remarks}
The main purpose of this paper has been the derivation of the exact
solution of the reduced equations which result from singular
perturbation methods. 
What has been shown is that the solution of the reduced equations up to
first order is equal to sum of the most divergent terms, which are
proportional to $\varepsilon t,\ \varepsilon^2t^2,\
\varepsilon^3t^3,\ldots$ appearing in the naive expansion.
In other words, taking up to only first order with respect to
perturbation parameter is enough to include those most divergent terms
in the approximate solution. 
The main result has been proved without any approximation.
Then it holds not only in the case $\varepsilon$ is small,
although this result is meaningful in the context of the perturbation
analysis.  

Another result has been presented in this paper.
That is a method to construct a perturbation solution
 where we make use of the Lie symmetry group which leaves the
system invariant.
With this method, we obtain recursive equations (\ref{eq:2-4}) instead
of Eqs. (\ref{eq:1-3}). 

For the future, it should be investigated that how the approximation
improves if higher order terms taken into consideration when we
construct the reduced equation.  
Another interest is the application to systems of partial differential
equations (PDE). 
In some PDE systems, it has been shown that, in constructing reduced
equations, we should take up not only most divergent terms in the naive
expansion but also other terms to preserve the symmetry of the original
system \cite{Graham1996}. 
Therefore, the proof presented in this paper should be modified properly
to those systems.

\begin{acknowledgments}
The author is grateful to Professor K. Nozaki, Professor T. Konishi
 T. Nishine, Nagoya University, H. Chiba, Kyoto University, and S. Goto,
 Lancaster University, for fruitful discussions. This research 
 is partially supported by a Grant-in-Aid from Nagoya University 21st
 Century COE (center of excellence) program ``ORIUM''.  
\end{acknowledgments}

\appendix
\section{The calculation of $E^i_{p_1p_2\ldots p_n}$ in Eq. (\ref{eq:3-5})}
At first, we consider the first term in the right hand side of
Eq. (\ref{eq:3-4}), 
$\phi^{(0)}_j\partial_{z_j}g^i$. 
Substituting Eq. (\ref{eq:3-1}) and Eq. (\ref{eq:3-3}) into
Eq. (\ref{eq:3-4}), and writing terms proportional to $t$, 
we obtain
\begin{eqnarray}
 \phi^{(0)}_j\partial_{z_j}g^i
  &=&\left(\sum_{\scriptsize{
               \begin{array}{c}
		p_1,p_2,\ldots,p_n\\
                \sum_{k=1}^n \lambda_kp_k-\lambda_j=0
	       \end{array}}
               }
           \hspace{-1.0cm}
           C^j_{p_1 p_2 \ldots p_n}t\prod_{k=1}^n z_k^{p_k}
   \right)
   \partial_{z_j} 
   \left(\sum_{\scriptsize{
                q_1,q_2,\ldots,q_n
                          }
              }
           C^i_{q_1 q_2 \ldots q_n}\prod_{k=1}^n z_k^{q_k}
   \right)
    \nonumber \\
  &&+ [{\rm terms\ not\ proportional\ to\ }t  ].
 \label{eq:A-1}
\end{eqnarray}
Those terms which are zero eigenfunctions of
$\sum_{k=0}^n\lambda_k z_k \partial_{z_k}-\lambda_i$ 
cause terms proportional to $t^2$ in $\phi^{(1)}_i$. 
Such terms in the right hand side satisfies the resonance condition:  
\begin{eqnarray}
 \sum_{\scriptsize{
         \begin{array}{c}
           k=1\\
           k\neq j
	 \end{array}}
       }
     ^n
  \lambda_k(p_k+q_k)+\lambda_j(p_j+(q_j-1))-\lambda_i=0,
 \label{eq:A-2}
\end{eqnarray} 
which reads
\begin{eqnarray}
  \sum_{k=1}^{n}\lambda_kq_k-\lambda_i=0
 \label{eq:A-3}
\end{eqnarray}
since $\sum_{k=1}^{n}\lambda_kp_k-\lambda_j=0$. 
Then, the resonant terms which is proportional to $t$ in the right hand
side of Eq. (\ref{eq:A-1}) become
\begin{eqnarray}
  t\left(\sum_{\scriptsize{
               \begin{array}{c}
		p_1,p_2,\ldots,p_n\\
                \sum_{k=1}^n \lambda_kp_k-\lambda_j=0
	       \end{array}}
               }
           \hspace{-1.0cm}
           C^j_{p_1 p_2 \ldots p_n}\prod_{k=1}^n z_k^{p_k}
   \right)
  \partial_{z_j} 
  \left(\sum_{\scriptsize{
               \begin{array}{c}
		q_1,q_2,\ldots,q_n\\
                \sum_{k=1}^n \lambda_kq_k-\lambda_i=0
	       \end{array}}
               }
           \hspace{-1.0cm}
           C^i_{q_1 q_2 \ldots q_n}\prod_{k=1}^n z_k^{q_k}
   \right)
 \label{eq:A-4}
\end{eqnarray}
On the other hand, for the second term in the right hand side of
Eq. (\ref{eq:3-5}), $g^j\partial_{z_j}\phi^{(0)}_i$,
\begin{eqnarray}
 g_j\partial_{z_j}\eta^{(0)}_i
  &=& \left(\sum_{\scriptsize{
                q_1,q_2,\ldots,q_n
                          }
              }
           C^j_{q_1 q_2 \ldots q_n}\prod_{k=1}^n z_k^{q_k}
   \right)
   \partial_{z_j} 
   \left(\sum_{\scriptsize{
               \begin{array}{c}
		p_1,p_2,\ldots,p_n\\
                \sum_{k=1}^n \lambda_kp_k-\lambda_i=0
	       \end{array}}
               }
           \hspace{-1.0cm}
           C^i_{p_1 p_2 \ldots p_n}t\prod_{k=1}^n z_k^{p_k}
   \right)
 \nonumber \\
 &&+ [{\rm terms\ not\ proportional\ to\ }t  ]
\label{eq:A-5}
\end{eqnarray}
The resonance condition in this case becomes
\begin{eqnarray}
 \sum_{\scriptsize{
         \begin{array}{c}
           k=1\\
           k\neq j
	 \end{array}}
       }
     ^n
  \lambda_k(p_k+q_k)+\lambda_j((p_j-1)+q_j)-\lambda_i=0,
 \label{eq:A-6}
\end{eqnarray}
which reads
\begin{eqnarray}
 \sum_{k=1}^{n}\lambda_kq_k-\lambda_j=0,
 \label{eq:A-7}
\end{eqnarray}
since $\sum_{k=1}^{n}\lambda_kp_k-\lambda_i=0$.
Then, the resonant terms which is proportional to $t$ in the right hand
side of Eq. (\ref{eq:A-5}) become
\begin{eqnarray}
  t\left(\sum_{\scriptsize{
               \begin{array}{c}
		q_1,q_2,\ldots,q_n\\
                \sum_{k=1}^n \lambda_kq_k-\lambda_j=0
	       \end{array}}
               }
           \hspace{-1.0cm}
           C^j_{q_1 q_2 \ldots q_n}\prod_{k=1}^n z_k^{q_k}
   \right)
   \partial_{z_j} 
  \left(\sum_{\scriptsize{
               \begin{array}{c}
		p_1,p_2,\ldots,p_n\\
                \sum_{k=1}^n \lambda_kp_k-\lambda_i=0
	       \end{array}}
               }
           \hspace{-1.0cm}
           C^i_{p_1 p_2 \ldots p_n}\prod_{k=1}^n z_k^{p_k}
   \right)
\end{eqnarray} 
This is equal to Eq. (\ref{eq:A-4}). 
Thus, it is shown that resonant terms which is proportional to $t$ in
$\phi^{(0)}_j\partial_{z_j}g_i-g_j\partial_{z_j} \phi^{(0)}_i$
for all $j$ and $i$ is equal to zero. 
That is to say, all of $\{E^i_{p_1p_2\ldots p_n}\}$ in
Eq. (\ref{eq:3-5}) is equal to zero.

\end{document}